\documentclass[twocolumn,10pt,cleanfoot]{asme2ej}

\usepackage{epsfig} 
\usepackage{amsmath}
\usepackage{comment} 
\usepackage{graphicx}
\usepackage{setspace}
\usepackage{graphicx}
\usepackage{mathtools}
\usepackage{placeins}
\usepackage{afterpage}
\usepackage{tocloft}

\title{A method to predict location of non-coup brain injuries
}

\author{
Rajesh Kumar${}^{a}$, Md Zubair ${}^{a}$, Sudipto Mukherjee${}^{a}$ and Jacobo Antona Makoshi${}^{b}$
\affiliation{${}^{a}$
              Department of Mechanical Engineering
              Indian Institute of Technology Delhi, India; \\ ${}^{b}$
              
              Japan Automobile Research Institute, Ibaraki, Japan \\
              *mez188285@iitd.ac.in
}}
\begin{document}
\maketitle
\begin{abstract}
{\it Brain injuries are a major reason for mortality and morbidity following trauma in sports, work and traffic. Apart from the trauma at the site of impact (coup injury), other regions of the brain remote from the impact locations (non-coup) are commonly affected. We show that a screw theory-based method can be used to account for the combined effect of head rotational and linear accelerations in causing brain injuries. A scalar measure obtained from the inner product of the motion screw and the impact screw is shown to be a predictor of the severity and the location of non-coup brain injuries under an impact. The predictions are consistent with head impact experiments conducted with non-human primates. The methodology is proved using finite element simulations and already published experimental results \\
$\mathbf{Keywords:}$ Non Coup Injuries, Screw Theory, Injury function, Primate Injuries}
\end{abstract}

\section{Introduction}
\label{intro}
Traumatic brain injuries (TBIs) are a major reason for mortality and morbidity in trauma \cite{thurman1999traumatic}\cite{hyder2007impact}. TBIs result from mechanical energy transferred to the brain from physical forces that act directly on the head or are transmitted through the head-neck complex \cite{three}. This mechanical energy produces deformations of the brain, its neurons, its supporting structures or its vasculature beyond tolerable levels which result in the injuries. Various methodologies including head impact experiments with animals have been used by researchers to investigate TBIs on impact \cite{five}. The head acceleration modes (rotational vs linear) and its magnitude have been historically related to the severity of brain injury \cite{five}\cite{six}\cite{forteen}. While it is expected that the brain regions in the surroundings of the head impact location will be affected, often there is one or more distal region of the brain that is injured (non-coup injury). Localisation of trauma within the brain has been researched considering the brain matter as a viscoelastic continuum \cite{seven}, but still remains a problematic issue. The skull motion (acceleration) is also used to predict brain injury using finite element (FE) model \cite{eight}\cite{nine}\cite{ten}\cite{eleven} by assessing strains in the brain tissue. Apart from the second derivative of motion, parameters like brain size and shape also affect the risk of injury \cite{twelve}. Consequently, multivariate regression (learning) techniques have also been used for vulnerability assessment of injuries \cite{thirteen}. A number of mathematical functions have been proposed to predict the likelihood of injuries during impact. One of such functions, is the Head Injury Criteria, has been adopted globally as a standard for motor vehicle and helmet safety evaluation.
The head injury criterion (HIC) uses linear acceleration parameters to determine the likelihood of the head injury during impact \cite{fifteen}. The HIC is given by a computable expression (equation \ref{HIC}).
\begin{equation}
HIC = max_{t_1,t_2}(t_2-t_1)(\frac{1}{t_2-t_1}\int_{t_1}^{t_2}a(t).dt)^{2.5}
\label{HIC}
\end{equation}

where, a(t) is the resultant acceleration measured at the centre of gravity of the head; $t_1$ and $t_2$ suggests the onset time and the end time of impact, respectively. The HIC has been criticized for not accounting for factors known to affect head injury, such as the impact direction, shape of the head, the area of contact or the rotational component of the head motion. Further, the HIC was developed on the basis of experimental data of severe head injury and it may not be applicable for milder forms of TBI, especially those not including a skull fracture \cite{39gennarelli1987directional}\cite{38got1983morphological}. 
Head impacts result in a combination of the linear and angular motion of the head. The probability of injury due to impacts that cause rotation free linear acceleration vs those caused by rotational motion (acceleration) have been examined \cite{sixteen} through helmeted head-form impacts and validated through injury outcomes of on-field football impacts. The contribution of angular rotation of the head on trauma, particularly on diffuse axonal injuries, has been demonstrated by experimenting on non-human primates \cite{40gennerelli1971comparison}\cite{41abel1978incidence}, rats \cite{43davidsson2009injury} and swine \cite{44fievisohn2014evaluation}. Comparable linear acceleration is also known to cause concussion and subarachnoid hematoma in primates \cite{36ono1980human}, rats \cite{45marmarou1994new} and swine \cite{44fievisohn2014evaluation}. Consequently, combinations of the linear and the rotational head motion have also been proposed as predictors of brain injury \cite{twentytwo}\cite{twentythree}. Further, regression models that relate the probability of concussion \cite{seventeen} with a combination of rotational and linear acceleration components have been developed, as illustrated by equation \ref{PIC}.
\begin{equation}
\textit{Probability of Concussion} = \frac{1}{1+e^-({\beta_0+\beta_1a+\beta_2\alpha + \beta_3a\alpha})}
\label{PIC}
\end{equation}

where $\beta_i$ are regression constants (where i = 0,1,2,3), ``a" is the peak linear acceleration and ``$\alpha$" is the peak rotational acceleration.
The effect of acceleration-time history of the head (post impact) \cite{eighteen} has also been studied in relation to brain injury \cite{ninteen}. Brian strain fields have been also assessed from experimental data on acceleration \cite{twenty}. FE methods are the only current method of identifying trauma locations and usually indicates regions in the periphery. Other methods predict existence of brain injury. We note that none of these studies incorporate information about the nature (direction) of incident impact.
An important aspect from the field data that is under-reported is the nature of impact that has induced the observed kinematics. The head kinematics resulting from the impact are related to the magnitude and direction of the impacting impulse. Specifically, the two are related by writing the Newton-Euler equations about the CG, or a point on the body fixed in space, or a point accelerating towards the CG. Impact to the brain, when measured externally, is available as a combination of angular acceleration and linear acceleration for the skull-bone with reference to a point of reference, which is usually the CG of the head. In contrast, the impact, when represented as a wrench \cite{twentyfive}\cite{30hunt1978kinematic}, is independent of the choice of the point of reference, as long as the observation is from a Newtonian frame of reference. Screw theory \cite{twentyfive} allows study of system dynamics of a rigid body as the effect of force motors (or wrenches) resulting in acceleration motors. Here the word ``motors” is indicative of a combination of a vector and a moment resulting from the vector. It also allows inferring the direction of impact knowing the location of impact, and the consequent head kinematics \cite{twentyfive}\cite{30hunt1978kinematic}. 
Protection by design of safety equipment, minimises the transmission of the impact energy to the brain. We hypothesize that the energy associated with small finite volumes of the brain due to the impact can be a measure of the dosage of the part. The instantaneous work done by the impact force at each mass node is representative of the stretching of segments the brain would undergo. To ascertain the critical regions, it is hence, sufficient to map an  energy quantifier at different positions when the body is impacted. The energy quantifier relates to the amount of energy transmitted to the soft part of the brain from the skull at different locations. 
The aim of our study is to formulate, leveraging screw theory, a function that relates a head impact wrench (force and moment combination) to the affected non-coup regions of the brain. The function developed utilises the direction of the impact to provide a metric of concussion. We support the theoretical formulations with examples of head impact experiments conducted on non-human primates in the 1970s, with numerical simulations of the past experiments conducted and published recently \cite{46antona2013correlation}\cite{47antona2012reanalysis}. The numerical values are compared with the standard metrics like HIC and the empirical formulation of probability of concussion.

\section{Methodology}
Non-coup injuries are those that occur at the distal location away from the impact. When the head sustains an impact, the trauma to a region of the brain is posited to be proportional to a quantification of energy that passes through the region. This is consistent with the product of strain $\epsilon$ and strain rate at the midbrain region, $\epsilon\frac{d\epsilon}{dt}_{max}$, providing one of the strongest correlation with the occurrence of mild traumatic brain injury \cite{twentyone}. In the conventional global head kinematics approach, the motion of the CG is studied as a representative point for assessment of injuries. The velocity (acceleration) and angular velocity (acceleration) of the CG (or any other point in the rigid body) are represented using six scalar variables. The theory of kinematic motors \cite{twentyfive} allows representation of the motion of rigid body as a whole using five scalar parameters. An inner product of motion variables to the impacting wrench with units compatible in SI units as Joules, which we call the energy is used to map the traumatic contour.
\subsection{Theoretical Framework}
The theory of kinematic screws is used to determine a map consisting of the instantaneous reciprocal product between the impacting wrench and the resulting screw. The function is defined across the brain volume by integrating the instantaneous working rate over the duration of the impact, to obtain a measure which we call as the Predictor Map. Maxima in the Predictor map gives the most probable region of non-coup injury, as it is a measure of energy passing through an affected region. The head has a reasonably rigid shell enclosing compliant matter. The energy associated is a simplification of the mechanics of the head. The different portions of the rigid part of the head interact with the soft part. The equivalent energy transmitted at each rigid-soft interface in the brain is considered to formulate an energy map for the non-coup injuries. For a given impact, the region with maximum energy transfer from the rigid to the soft part would be the portion of the brain leading with maximum non coup concussion.
\subsubsection{Idealised Geometry of Brain}
\begin{figure}
\centering

  \includegraphics[scale = 1]{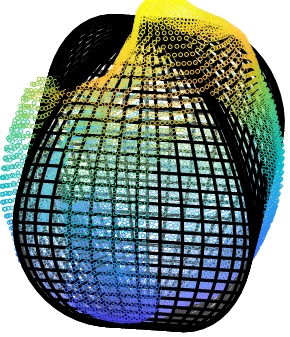}
\caption{Super-ellipsoid model developed as a simplistic representative of the primate (Macaque) brain}
\label{fig:1}       
\end{figure}

The brain has an irregular shape. However, geometrically simpler models of the brain, (ellipsoid, spheroidal or even spherical) have been traditionally used to develop theories by researchers \cite{26khalil1977parametric}\cite{27kenner1972dynamic}\cite{28chan1974mathematical}. In the current study, we, too adopt a super-ellipsoid as a simplified geometry representative of a primate brain (Figure \ref{fig:1}). This allows the function described to be calculated analytically.

\begin{figure*}
\centering

  \includegraphics[scale = 0.5]{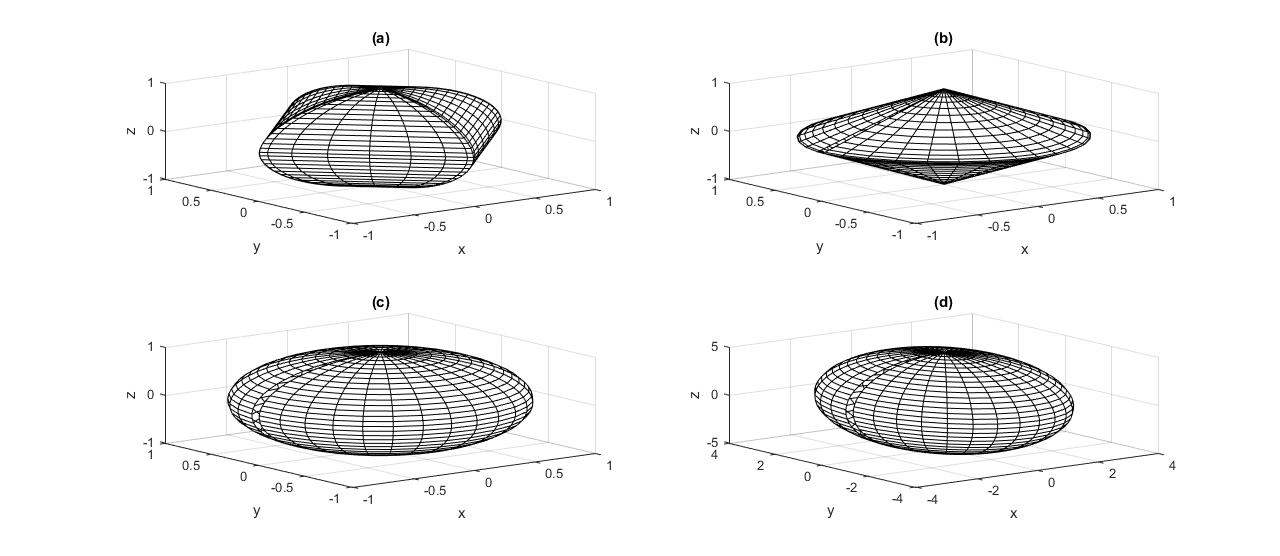}
\caption{Super ellipsoids with varying parameters (a) Equidistant diameters with h= 1.8, d = 1 used as the model for the brain. (b) Parameters: h = 1, d = 1.8, the model goes flatter moving away from a cube. (c) Parameters: h = 2, d=2, the output system is a sphere as the diameters along different axis are equal (d) Parameters: h = 2, d = 2 but with varying diameters along varying axes, forming an Ellipsoid.}
\label{fig:2}       
\end{figure*}

Throughout the paper, we use terminologies introduced in \cite{twentyfive} and \cite{30hunt1978kinematic}. The impact force is modelled as a wrench with zero moment component and with the axis same as that of the direction of the force. The brain geometry is assumed to be super ellipsoid as such geometries with variable parameters define surface transition from a cube to a sphere as plotted in Figure \ref{fig:2}. The super ellipsoids give sufficient form adaptability (equation \ref{supere}), and its first and second moments can be found in closed form. 
\begin{equation}
(|x/a_m|^d+|y/b|^d)^{\frac{h}{d}}+|z/c|^d \leq 1
    \label{supere}
\end{equation}
where x,y,z represents the cartesian coordinates, and $a_m, b, c$ are the parameters defining the symmetry of the geometry (for instance, it defines the magnitude of axes in a an ellipsoid) and h and d represent the shape of the system.  

A parametric formulation of the super ellipsoids is given in equation \ref{supere3} to \ref{supere4}.
\begin{equation}
    x(\delta,\gamma) = \rho{a_m} f_1(\gamma,2/h)f_2(\gamma,2/d)
    \label{supere3}
\end{equation}
\begin{equation}
     y(\delta,\gamma) = \rho{b} f_1(\gamma,2/h)f_2(\gamma,2/d)
\end{equation}
\begin{equation}
    z(\delta,\gamma) = \rho{c} f_2(\gamma,2/h)
\end{equation}
$\vee -\frac{\pi}{2}<\gamma<\frac{\pi}{2} and -\pi<\gamma<\pi$
\begin{equation}
    f_1(\eta,\zeta) = sgn(cos(\eta))|cos(\eta)|^{\zeta}
\end{equation}
\begin{equation}
    f_2(\eta,\zeta) = sgn(sin(\eta))|sin(\eta)|^{\zeta}
    \label{supere4}
\end{equation}
where $\gamma$ and $\delta$ are the parameters and $sgn(f_3)$ provides the sign of the output of $f_3$. For the super-ellipsoid, the moment of order ‘p,q,s’ with respect to x,y,z axes is defined as a Riemann integral \cite{32jaklic2003moments} (equation \ref{moment1}) and a closed form expression (equations \ref{moment2}-\ref{moment3}) allows direct computation of the moments.
\begin{equation}
M_{pqu} = \int_{-\infty}^{\infty}\int_{-\infty}^{\infty}\int_{-\infty}^{\infty}x^py^qz^uf(x,y,z)dxdydz
\label{moment1}
\end{equation}

\begin{equation}
\begin{split}
M_{pqu} = \frac{2}{p+q+2}a_m^{p+1}b^{q+1}c^{u+1}\epsilon_1\epsilon_2\\ \beta_f((u+1)\frac{\epsilon_1}{2},(p+q+2)\frac{\epsilon_2}{2}+1)\\\beta_f((q+1)\epsilon_1/2,(p+1)\epsilon_2/2+1)
    \end{split}
    \label{moment2}
\end{equation}

\begin{equation}
\epsilon_1 = \frac{2}{h}
\end{equation}

\begin{equation}
    \epsilon_2 = \frac{2}{d}
    \label{moment3}
\end{equation}

where $\beta_f$ is the standard beta function, given by $\beta_f(n_p,m_p) = \int_0^1t^{n_p-1}(1-t)^{m_p-1}dt$.\\
The external surface is discretised into a grid of rectangular patches. For each grid element the centroid of the rectangular patch is the point of reference for the element. We will subsequently reduce the motors (combination of vector and moments) to the points of reference of the elements to compute the wrenches needed to calculate the Predictor Map.
\subsubsection{Predictor Map Computation}
To quantify the function on a finite volume on the brain surface, we use the definition of instantaneous work done as the scalar product of the force and displacement at a point added to the product of the applied moment with angular motion. The reciprocal product \cite{31kim2003analytic} of an applied wrench with the resultant twist is a compact descriptor of the instantaneous work done. Conventionally, the reciprocity relation is converted to an inner product by reordering the vectors. The reciprocal product of a wrench and a twist screw is the sum of the vector inner product of the force component to linear velocity and vector inner product of the moment component with angular velocity.  
Following screw theory \cite{twentyfive} a set of forces ($\textbf{F}$) and moments ($\textbf{M}$) acting on a rigid body can be reduced to a wrench motor at the centroid of each grid element (equation \ref{dyn}) where $\mathbf{r}_{OJ}$ represents the radius vector from the centroid of the $j_{th}$ element to the point of application of the impacting wrench and $\mathbf{W}$ = $(\mathbf{F}$  $\mathbf{M})^{T}$ is the impacting wrench to the system. 
\begin{equation}
    \mathbf{R^j} =  \left(\begin{array}{cc}
        \mathbf{F} \\
         \mathbf{r_{OJ}}\times \mathbf{F} + \mathbf{M}
          \end{array}\right)
\label{dyn}
\end{equation}

Analogously, the instantaneous motion of a rigid body can be reduced to a twist motor of the type (equation \ref{kin})
\begin{equation}
     \mathbf{V^G_i} =  \left(\begin{array}{cc}
        \mathbf{v_i^G} \\
         \mathbf{r_{OJ}}\times \boldsymbol{\omega}^G
    \end{array}\right)
    \label{kin}
\end{equation}
where $\textbf{v}_i^G$ is the velocity vector of the centroid of the $i^{th}$ grid element, $\boldsymbol{\omega}^G$ is the angular velocity component and {G} is the frame fixed to the ground. In impact problems, the net motion of the body over the impact duration is usually small while velocity change can be substantial. It is hence reasonable to assume that the inertia binor (see appendix) of the system does not change significantly over the impact duration. For super-ellipsoids which are not excessively prolate, the differential of the binor of inertia with respect to time can hence be disregarded giving us the following equation relating wrench and twist (equation \ref{dyneq})
\begin{equation}
\mathbf{T}\frac{d\mathbf{V}}{dt} + \mathbf{V}\frac{d\mathbf{T}}{dt} = \mathbf{W}
    \label{dyneq}
\end{equation}
where, \textbf{T} is the inertia binor and \textbf{W} is the impacting wrench. Development of dynamics based on screw theory with definitions of the parameters of the dynamics’ equation is presented in the Appendix.
Let the displacement motor of the centroid of the grid element J be $\mathbf{\phi_J}$. The work done, $E^j$, by the impacting wrench affecting element ‘j’ $(R^j)$ to displace the considered grid element by $\boldsymbol{\phi_j}$ is $E^j$= $<\mathbf{R^j},$ $\boldsymbol{\phi_j}>$ which is a scalar quantity. The displacement motor is the component wise integral of the twist screw and the reciprocal product is the inner product of the element reduced wrench and displacement motor.
\subsection{Concept Evaluation - Methodology and Examples}
To verify the methodology proposed to predict the major regions of concussion, we used data from impacts controlled on a group of primates. This methodology is presented in section \ref{markit}. The results are also compared with the FE simulation results. 
\subsubsection{Experiments on Primates}
\label{markit}
\begin{figure*}
\centering

  \includegraphics[scale = 1]{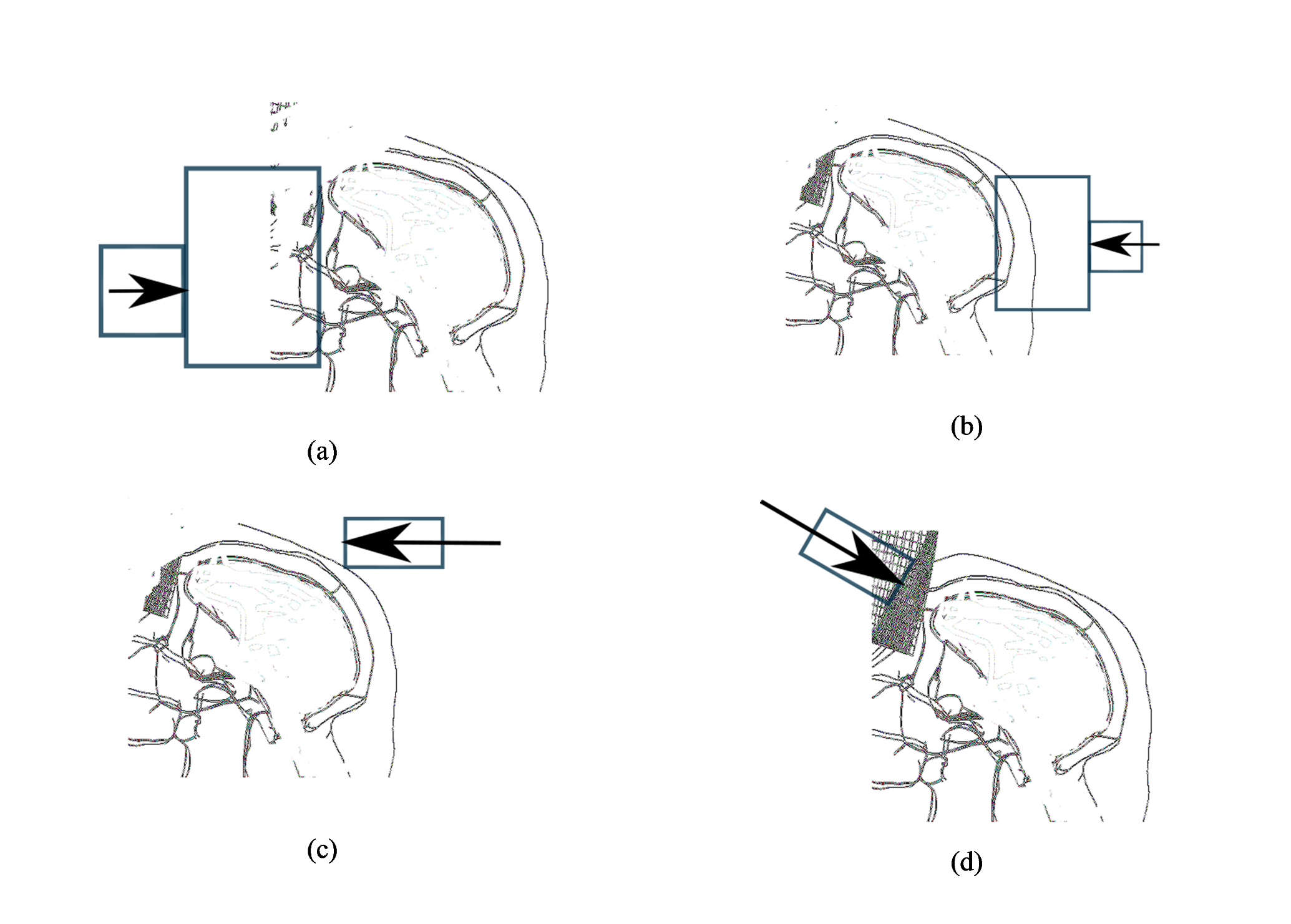}
\caption{(a) A padded normal dominant impact from front. (b) A padded normal dominant impact from rear (c) A dominant tangential impact (d) A combination of a normal and tangential impact.}
\label{testcase}       
\end{figure*}

Series of head impact experiments in controlled environments with non-human primates were conducted at the Japan Automobile Research Institute (JARI) in the 1970’s \cite{33masuzawa1976experimental}. The impacts can be grouped into three classes, those in which the impact was nearly normal to the frontal region, those in which the impact was tangential to the top surface of the skull and impacts that were off normal to impacted surfaces. These tests produced injuries of severity that were not consistent with HIC as a measure and structural damage appeared in unpredictable zones. We shall demonstrate that the approach proposed resolves this anomaly. The impacting wrench, for the cases where it was not explicitly recorded, was calculated from the acceleration data (equation \ref{dyneq}). Thereafter, the experimental data was classified into either the impact being near normal or near tangential or intermediate. As the impacting velocity in all cases was similar, wrench of same normalized magnitude was used in our analysis. The impacting force directions for various test clusters are shown in Figure \ref{testcase}. The dominant normal impacts are realized using a padded impactor. The experimental data has been published over the years in 1980s by JARI \cite{33masuzawa1976experimental}\cite{35sekino1980experimental}\cite{36ono1980human}\cite{37sakai1982experimental}.

\section{Results}
\label{res}
 To demonstrate the approach developed in section 2.1, a time varying normalised ramp impulse has been applied to various locations of the head model and the parts of the brain with top 10 \% of the energy function value are tracked to analyse the most critical areas under impact. This approach throws new light on the possible location of non-coup injuries and its dependence on the nature of the initial impact, parameterised in terms of direction and magnitude. Trauma locations from extensive experiments done in the 1970’s on monkeys at JARI, Japan, is shown to be consistent with this methodology. The finite element simulation results published by the co-author of this paper in \cite{46antona2013correlation} are also shown to be consistent with this methodology. We, hence, show that by combining with the direction of impact rather than using only the overall head kinematics, good injury predictability, in magnitude and location is achieved.
 \subsection{Simulation based on the theoretical framework}
\begin{figure}[h]
\centering
\includegraphics[scale = 0.6]{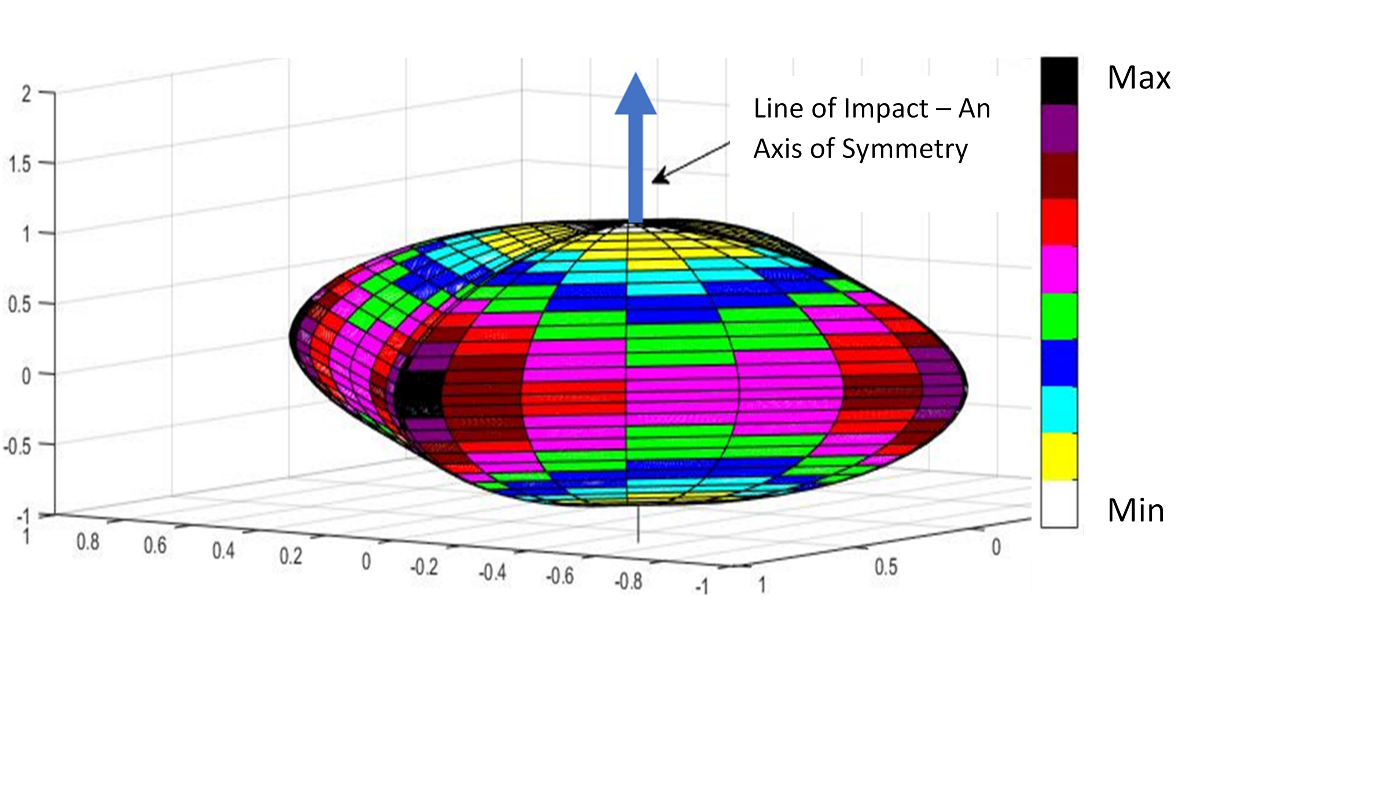}
\caption{For the impact near poles, the maximum effect is split along the four pseudo edges of the super ellipsoid. The dark regions suggest regions of traumatic injury}
\label{normaltop}       
\end{figure}

\begin{figure*}[h]
\centering

\includegraphics[scale = 1]{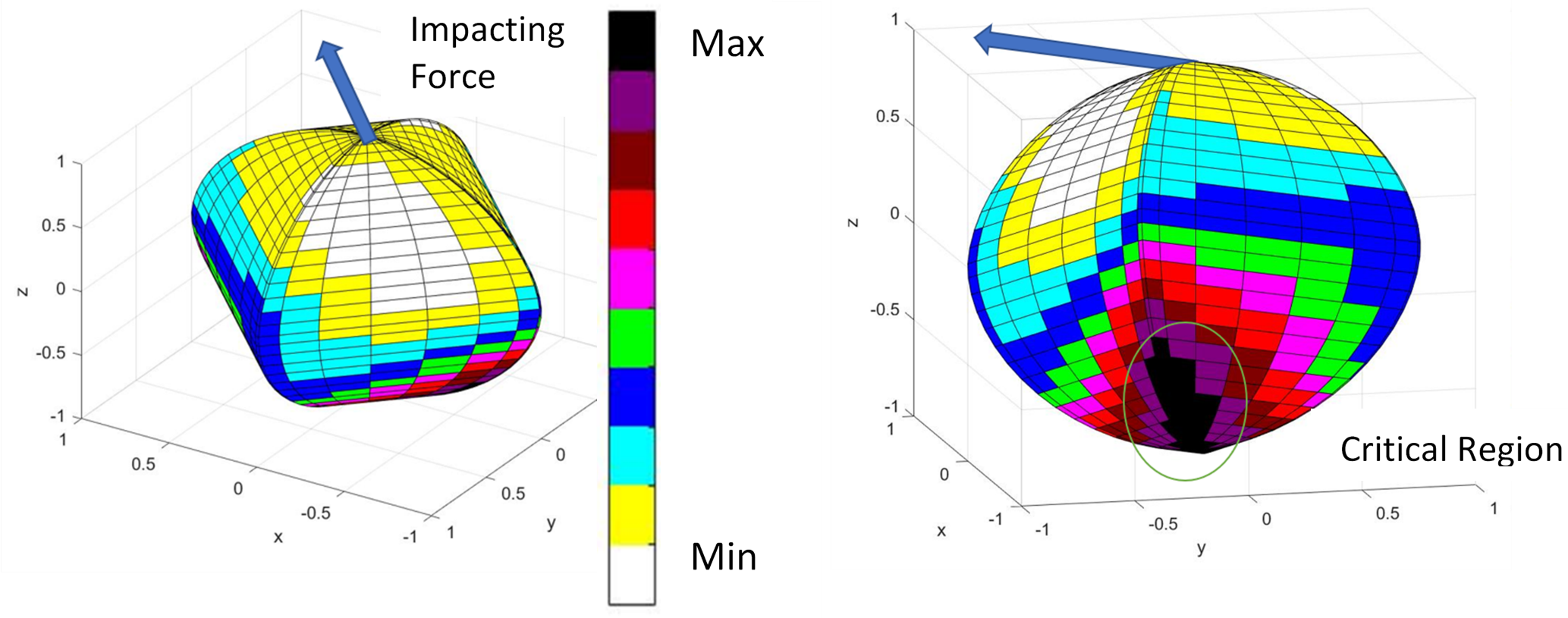}
\caption{Response of the function developed under tangential impact at the pole. (a) and (b) shows two perspective views of the response of the brain under impact. The arrow shows the direction of the impact.}
\label{tangent}       
\end{figure*}

\begin{figure}[h]
\centering

\includegraphics[scale = 0.2]{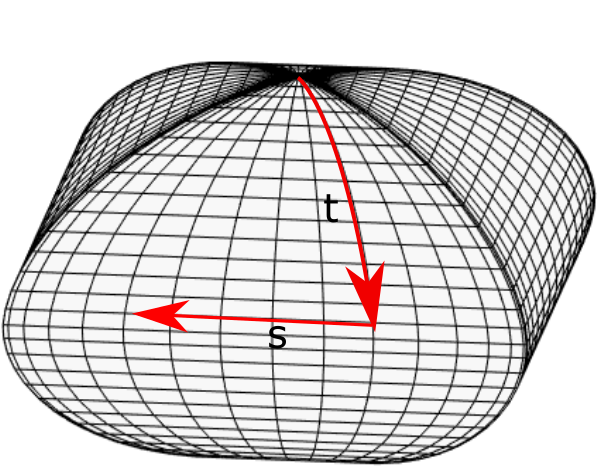}
\caption{Response of the function developed under tangential impact at the pole. (a) and (b) shows two perspective views of the response of the brain under impact. The arrow shows the direction of the impact.}
\label{definition}       
\end{figure}

\begin{figure}[h]
\centering

\includegraphics[scale = 0.18]{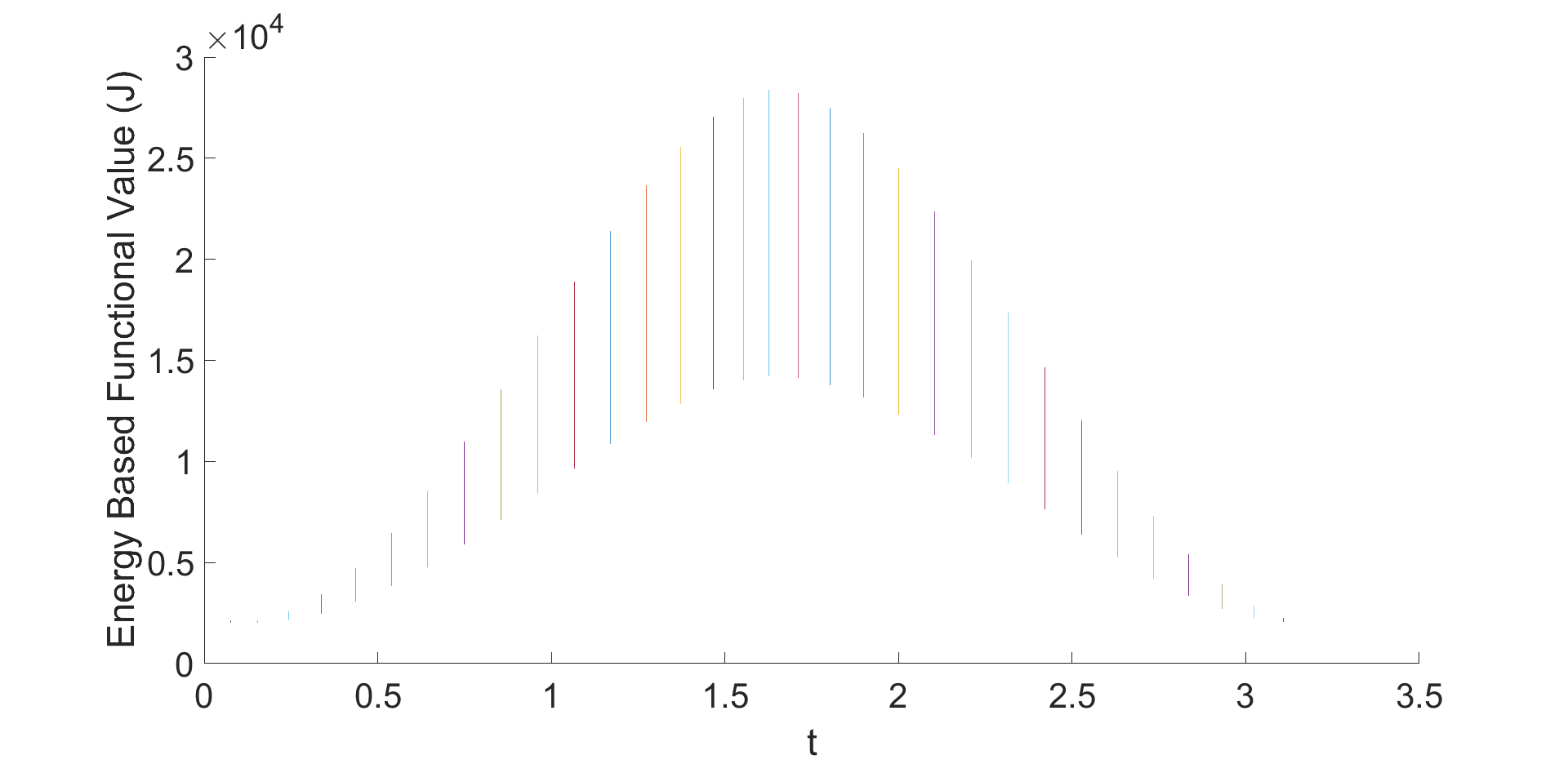}
\caption{Variation of the energy predictor function with respect to the variation of the longitudinal parameter for near normal impact. The length of each segment for a “t” value is an indicator of the variation with the “s” value}
\label{7variation}       
\end{figure}

\begin{figure}[h]
\centering

\includegraphics[scale = 0.2]{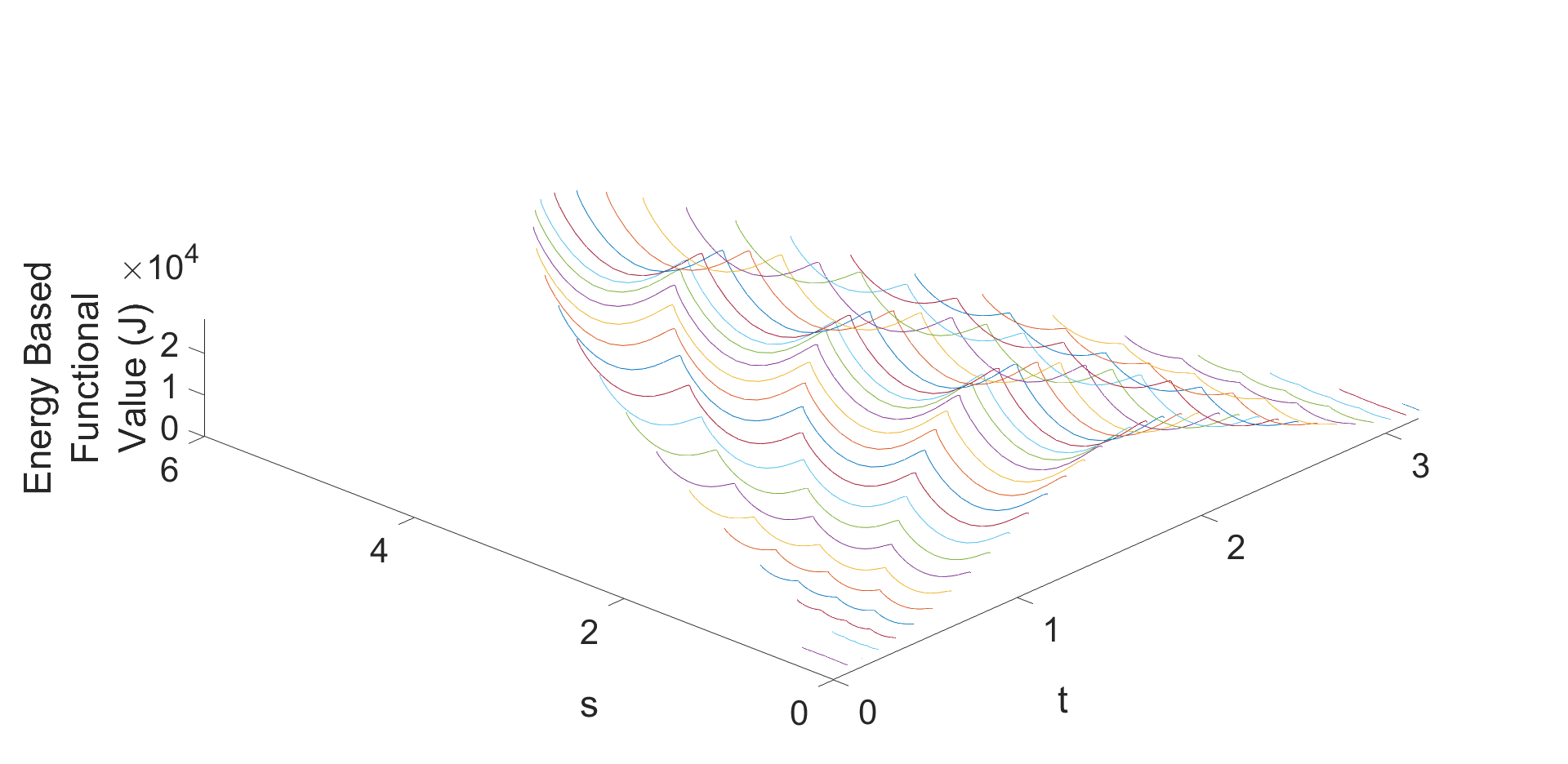}
\caption{Variation of the energy predictor function in the `s-t' domain for near normal impact}
\label{8variation}       
\end{figure}

 Figure \ref{normaltop} shows the case of an impact on the pole of the super ellipsoid. The arrow shows the direction of impact and pole of the arrow is the point of impact. The line of impact passes through the lower pole as well as the upper pole, making it along an axis of symmetry. Different values of the predictor map are colour coded and indicated by the bar on the right side of the figure, with darker colors indicating higher levels. The higher-level elements are concentrated along the four pseudo edges of the volume. The most traumatic region for the normal impact is at the farthest possible point in the volume from the line of impact. As the impact is along the axes of symmetry all the four regions are approximately at the same distance from the line of impact, which are all approximately maximum.
Figure \ref{tangent} shows the predictor map obtained for an impact at the same location, but tangential direction to the surface. This case represents on idealization of gliding type impacts and illustrates the change of the locations of peak value elements on modifying the impact direction.
For the super-ellipsoid geometry, two path parameters (t,s) (latitudinal and longitudinal) are naturally defined (Figure \ref{definition}). The contours showing the function value of the predictor function along the path parameters are extracted to study the location of the critical regions from the point of impact. The contour plots, parameterized by ``s" and ``t" are conveniently mapped to 2D, which is amenable to a graphical interpretation.
For a near-normal impact, the variation of the function value with respect to the longitudinal parameter, `t', is shown in Figure \ref{7variation}. The bars represent the variation with `s' and we see a wavelike change with `t'. The `t-s' surface is shown in Figure \ref{8variation}, which shows an additional periodic variation of the function in `s'. For near normal impact, the traumatic region is on the sides (central longitudinal parameter). Generally, for near normal impact, the most critical region of injury (apart from just at the point of impact) is at the point of farthest distance from the line of the impacting force. Near-normal impact means the impact direction passing through a bounded region in the vicinity of the geometric centre of the geometry. Similarly, near tangential impact is nearly perpendicular to the near normal impact direction. 

\begin{figure}[h]
\centering
\includegraphics[scale = 0.18]{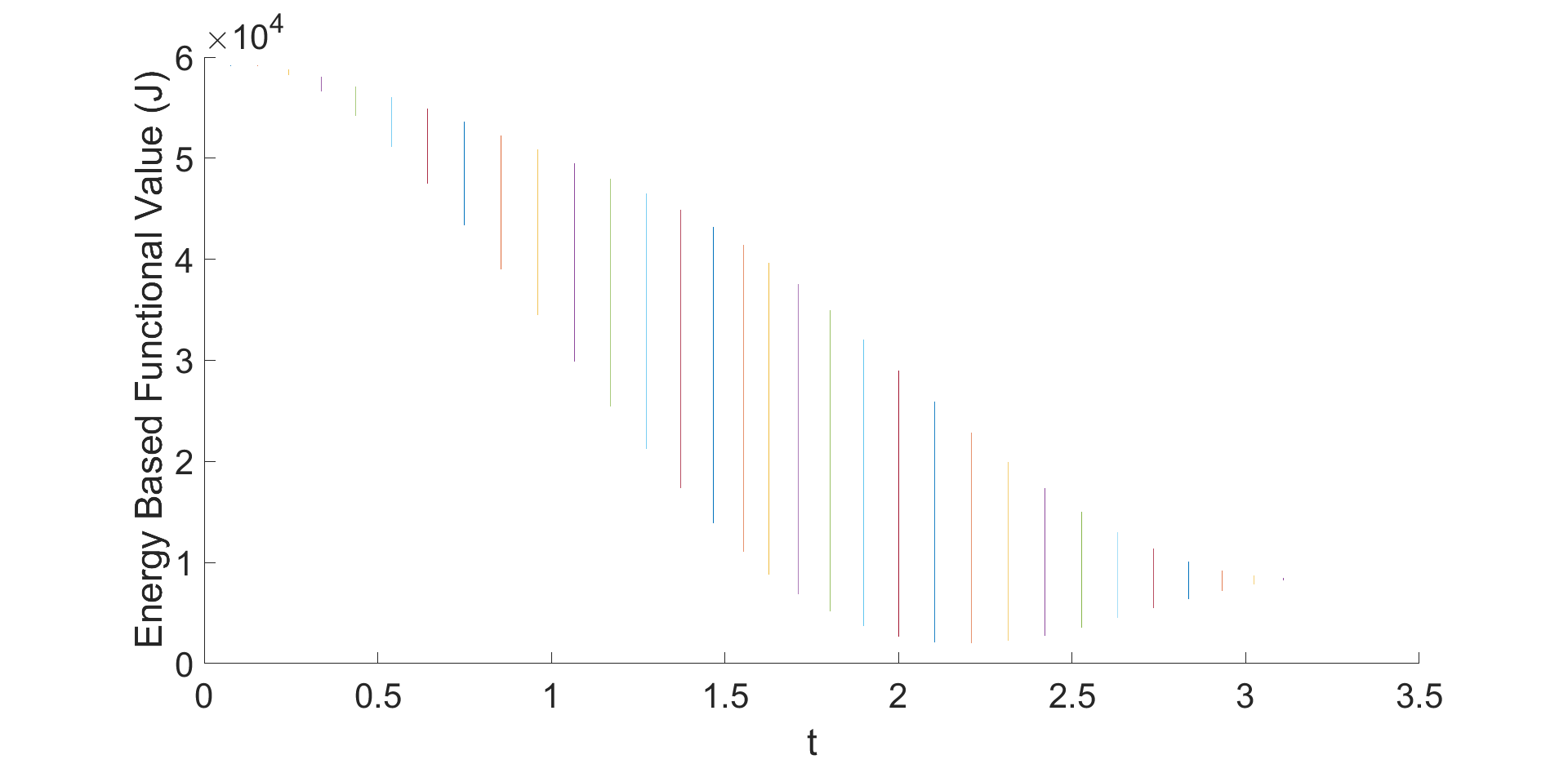}
\caption{Variation of the Predictor function with the longitudinal parameter for tangential impact.}
\label{9variation}       
\end{figure}

\begin{figure}[h]
\centering
\includegraphics[scale = 0.18]{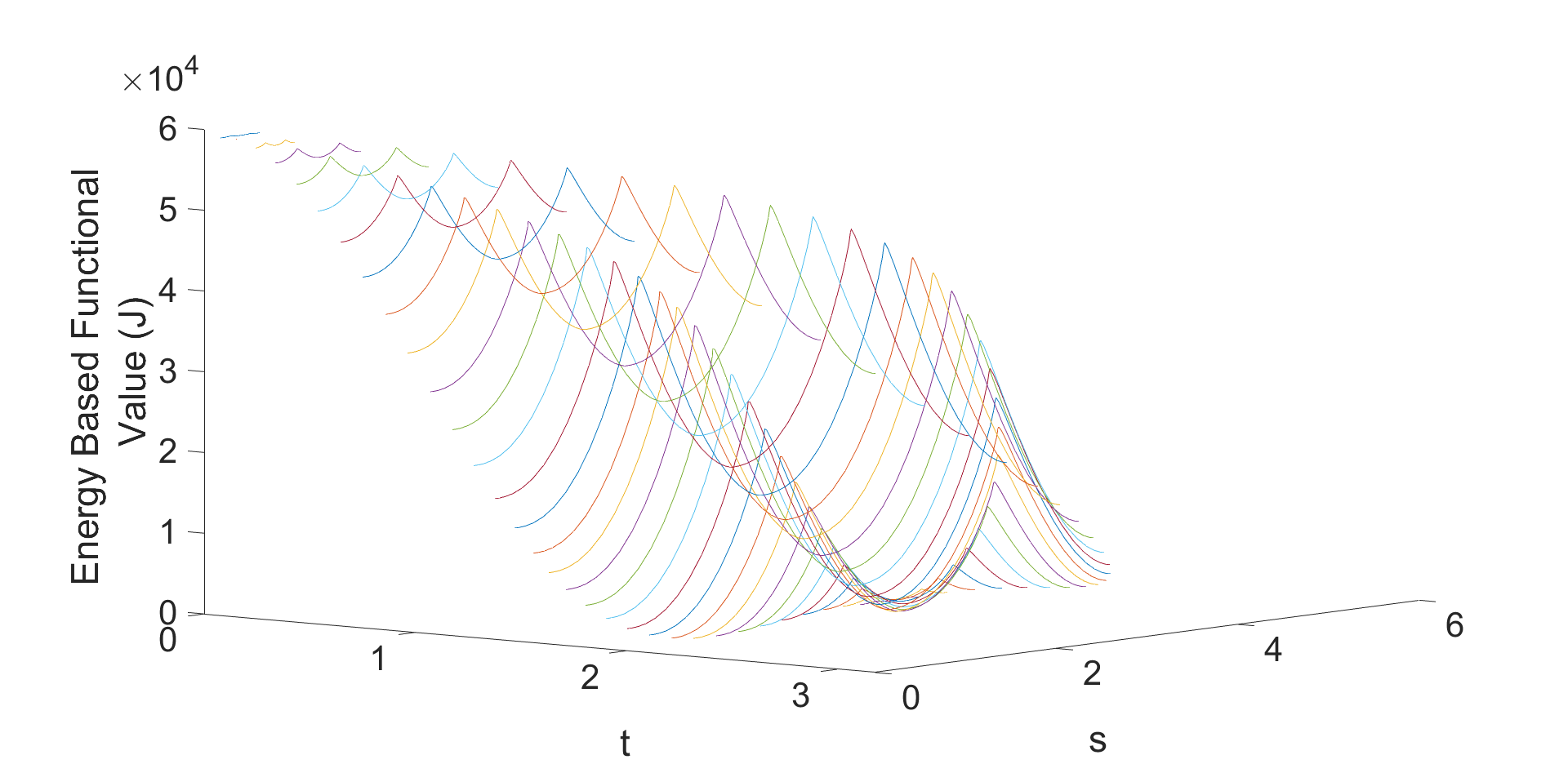}
\caption{Formation of the energy based function surface for tangential impact.}
\label{10variation}       
\end{figure}

In the figures above, the impact is at the poles. A force of magnitude 5000 N is applied at the pole for an instance of 0.02 seconds. The normal impact leads to bifurcation in the traumatic zones into four regions that are equally dispersed with respect to the central axes. However, in case of tangential impact at the pole, the maximum trauma is located at the maximum distance from the point of impact, which is the opposite side pole. The effect of tangential impact is shown in Figure \ref{9variation} and Figure \ref{10variation}. 

\begin{figure}
\centering
\includegraphics[scale = 0.5]{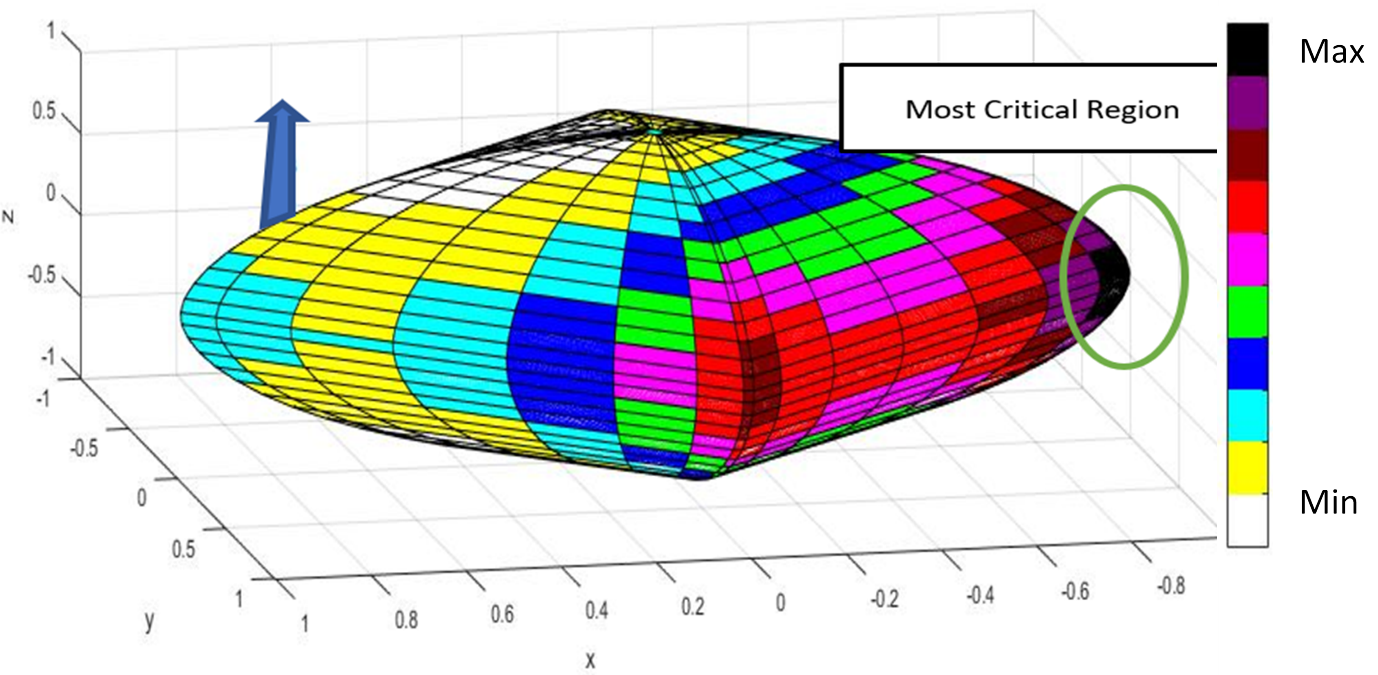}
\caption{Most critical region for a tangential impact is farthest from the point of impact. }
\label{11}       
\end{figure}

An additional observation is that when the predictor map for tangential impact as the normal impact of the same magnitude, the trauamatic zone is more localised. It would be interesting to conjecture that this would lead to more trauma than normal impact where the maximum area region is split in the `s' domain. The normalised maximum predicted energy is of around 27500 J, where as it is around 58000 J for tangential impact (Figure \ref{8variation} and Figure \ref{10variation}). When the model is tangentially impacted at the super ellipsoid surface (near the side), the most critical regions are again farthest from the point of impact (the rear of the head) at a single location. The two views showing the most critical region for this are shown in Figure \ref{11} where the arrows are representative of the impact direction. The tail of the arrow is the point of impact. In Figure \ref{11}, the impact point is nearly at the centre of the face. The darkest regions are located farthest from the point of impact.
\begin{figure}[h]
\centering
\includegraphics[scale = 0.25]{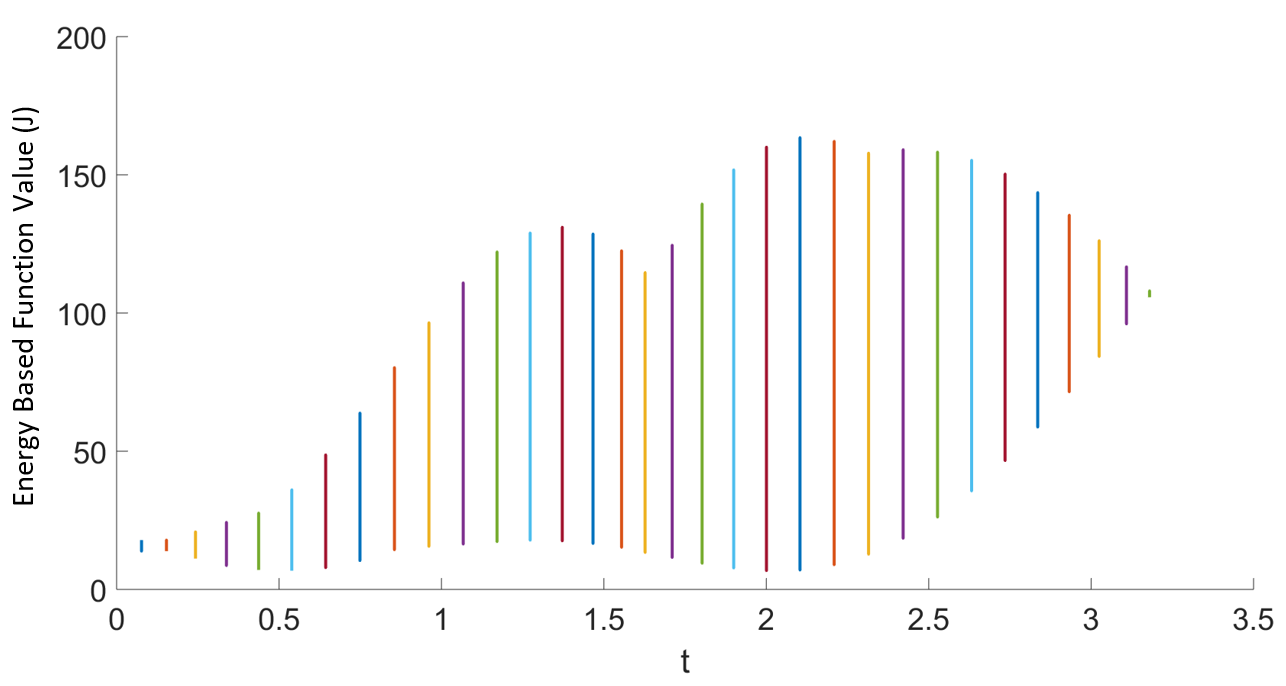}
\caption{Binormal Function Value region for general impacting wrench}
\label{fig12}       
\end{figure}
\begin{figure}[h!]
\centering
\includegraphics[scale = 0.2]{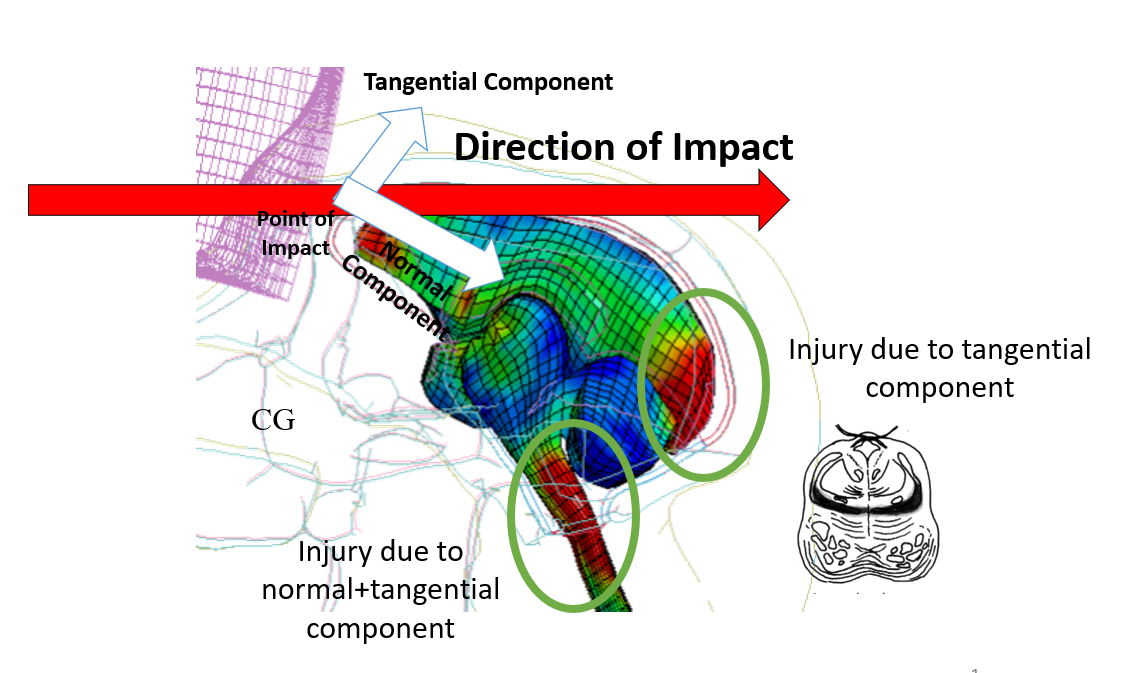}
\caption{Comparison of the finite element simulation results with the methodology developed.}
\label{fig_jacobo14}       
\end{figure}
\begin{figure*}[h]
\centering
\includegraphics[trim={3cm 7cm 0cm 0cm}, clip, scale = 1]{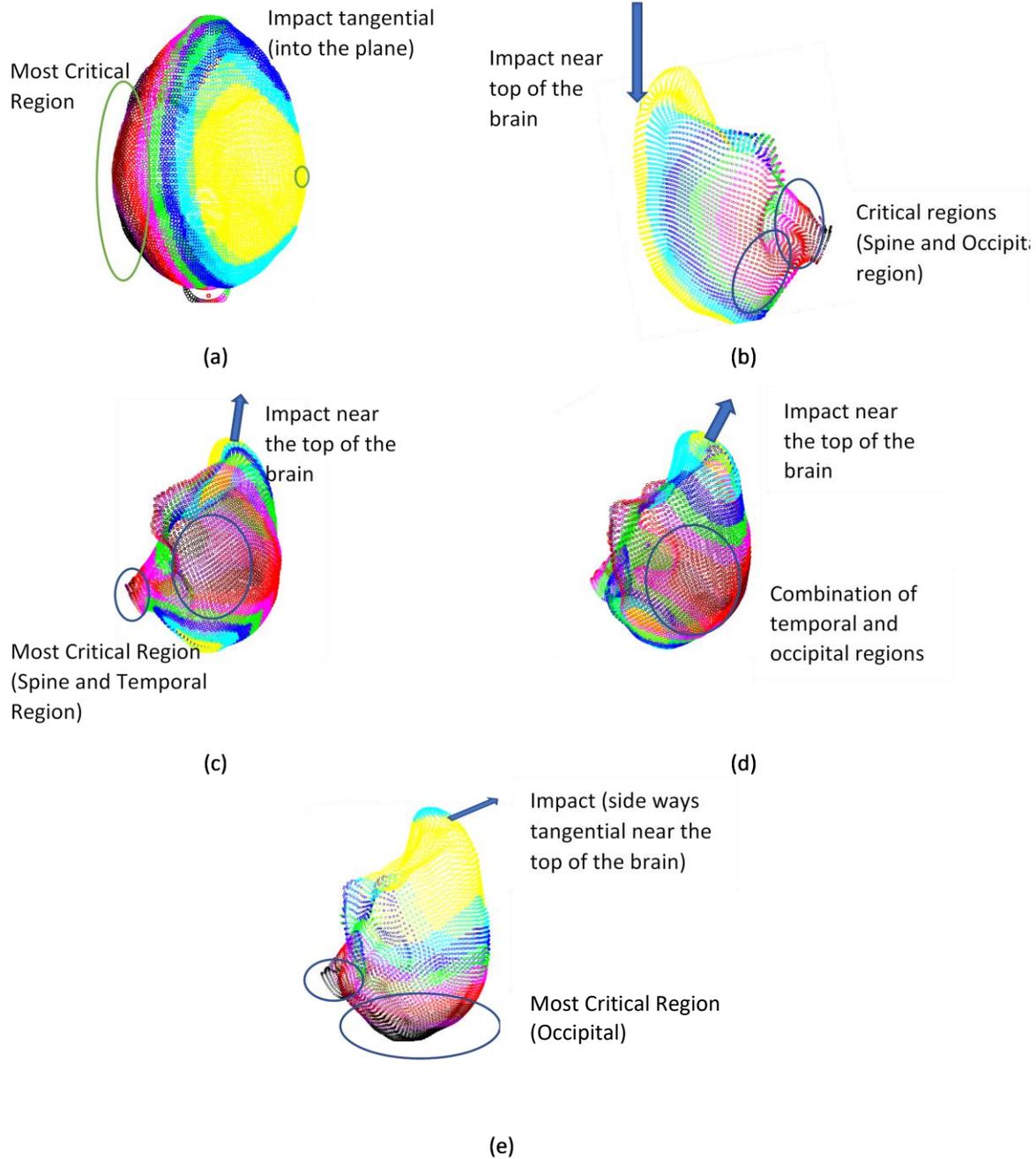}
\caption{Simulation on Primates' Brain for different impact locations and direction}
\label{comp_monkey15}       
\end{figure*}
\begin{figure*}[h]
\centering
\includegraphics[scale = 0.2]{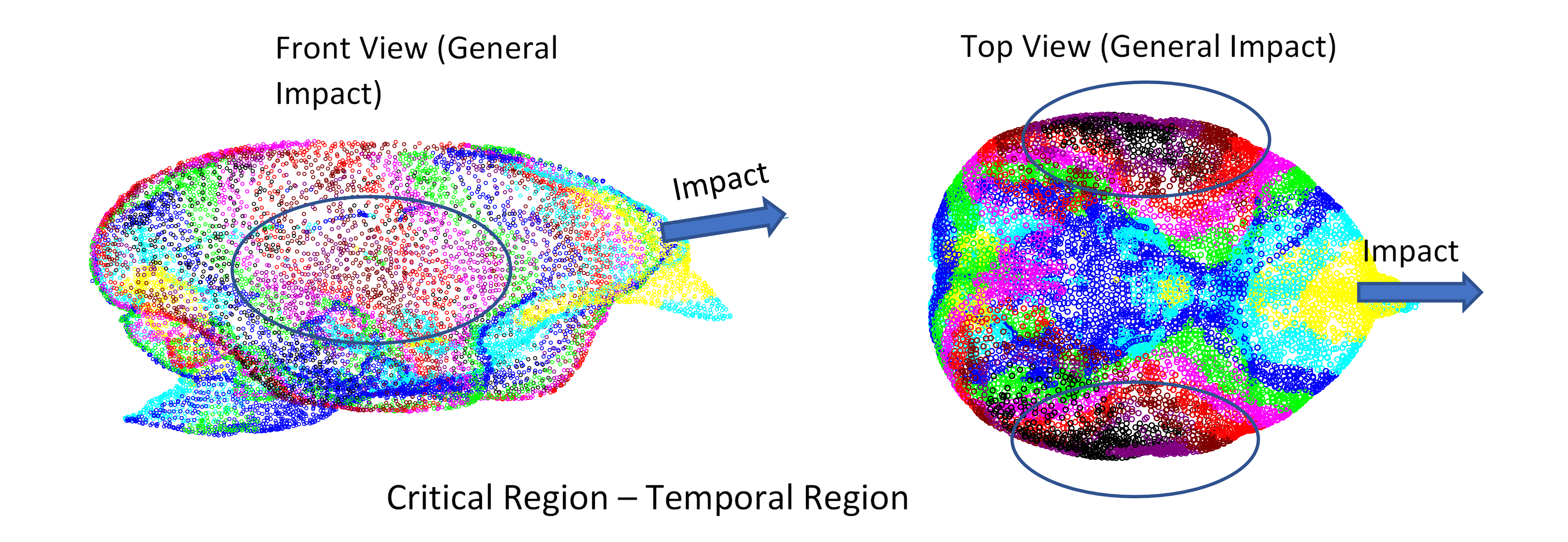}
\caption{Simulation on Marmoset Brain for a general impact direction (two views presented)}
\label{marmoset}       
\end{figure*}
\begin{figure}[h]
    \centering
    \includegraphics[trim={4cm 18cm 4cm 2cm},clip,scale = 0.6]{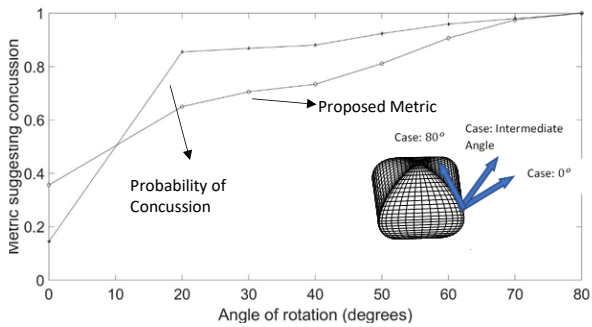}
    \caption{Comparison of Probability of Concussion with proposed Metric. The value of the metric proposed here was normalised by the peak value (80662 J) for proper comparison}
    \label{show}
\end{figure}
\begin{figure*}[h]
    \centering
    \includegraphics[scale = 1, trim={2cm 20cm 2cm 2cm}, clip]{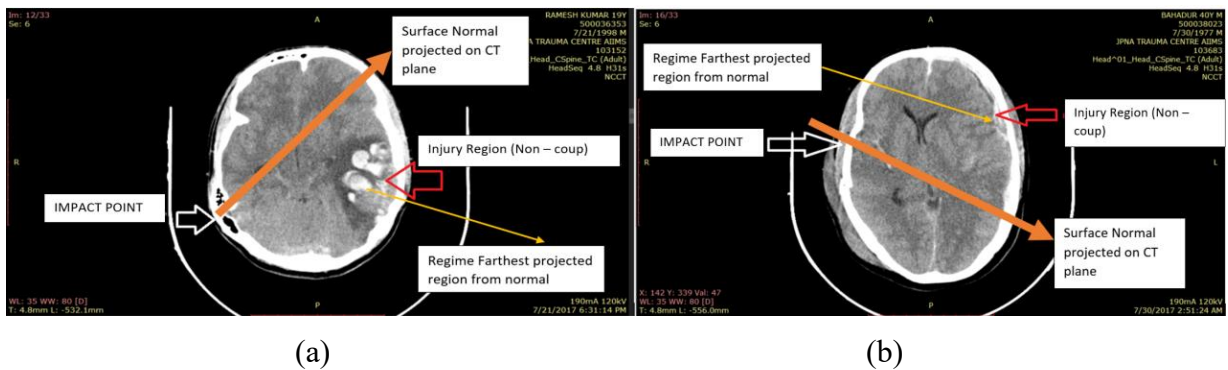}
    \caption{CT Image of a patient suffering head injury. Two cases of different patients are presented in (a) and (b). In both the cases the injury region is farthest from the projected normal to the surface of the brain geometry. The normal drawn are approximately projected normal to the CT plane}
    \label{fig:17}
\end{figure*}
\begin{figure}[h]
    \centering
    \includegraphics[scale = 0.6, trim={2cm 20cm 2cm 2cm}, clip]{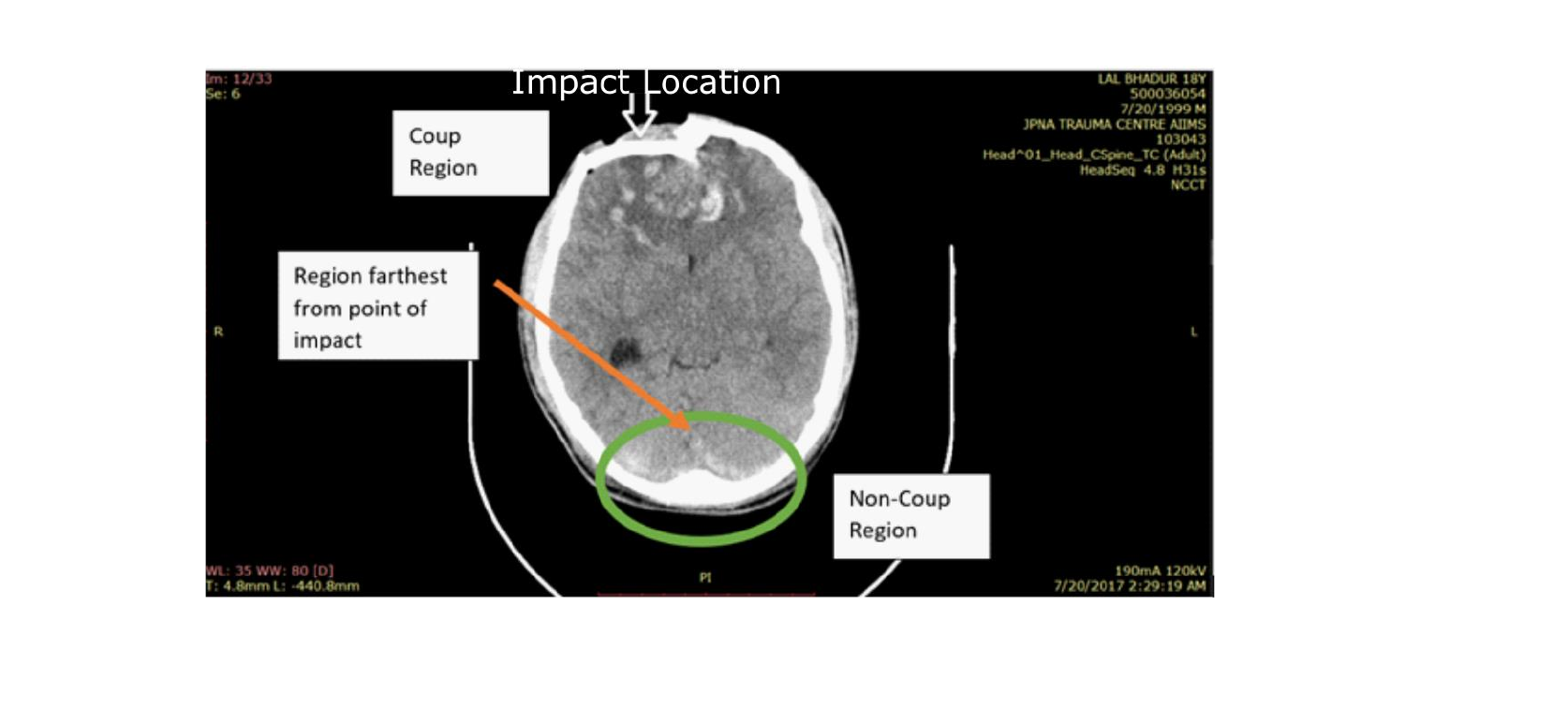}
    \caption{CT Image of a patient suffering head injury via frontal impact}
    \label{fig:18}
\end{figure}
For a general impact, with components, along both normal and tangential direction, the resulting contour is bi-normal in the `t' domain, resulting from the combination of the two modes and is shown in Figure \ref{fig12}. The resulting contour can be seen as a combination of the two modes presented above.
In summary, in the cases of near tangential impact, the critical regions are the areas farthest from the point of impact. The critical zones, in this case does not change rapidly with the direction of impact. Whereas in case of near normal impacts, the critical regions are farthest from the line of impact, and hence dependent on the impact direction. For the general impact force which cannot be classified as normally dominated or tangentially dominated impact, the resultant areas of maximum injury are a combination of the two canonical modes. The zone with higher Predictor value or the traumatic region is focused with larger magnitude in case of a tangential impact.

\subsection{Comparison with Experimental Results on Primates}
To analyse the experimental data, we have classified the cases based on the direction of impact with respect of the surface normal. Section 3.2.1 compares the near normal impact to the simulation results, section 3.2.2 deals with near tangential impacts whereas section 3.2.3 is based on general impacts which are a combination of near normal and near tangential impacts. 
\subsubsection{Near Normal Impacts – Comparison with Primates’ results}
The experimental results of the near normal impacts, were reported in \cite{33masuzawa1976experimental}. The impact is on the frontal region of the skull and nearly normal to the surface. Literature suggests that the contra-coup zones \cite{34goggio1941mechanism} are expected on the surface diametrically opposite to the impacted surface, due to wave transmission. But several cases in the series report anomalous injuries in other regions as well. We show next that they are in zones predicted by the model proposed here.

Upon experimentation on primates by JARI, the experimental results of frontal impacts on 13 primates were published \cite{33masuzawa1976experimental}. Out of 13 impacts, five of them showed no residual injury and four of them had injuries at regions apart from the impacted regions. The macroscopic findings for the non-coup injuries included injuries near the base of the brain geometry. Additionally, in two of the cases, injuries were also present in the ``occipital" and ``temporal" lobe. For the cases with no macroscopic findings of the injured region, microscopic injuries were notified at the ``brain stem" including the effect on the brain stem neurons. In all cases the primate experimental results indicate that the effect of normal impact results in injuries around the region farthest from the line of impact. The same is predicted by the mechanics-based model.
\subsubsection{Near Tangential Impacts - Comparison with Primate results}
The case of primates being impacted near the top of the head tangentially, were reported in \cite{35sekino1980experimental}. The report suggested that 7 out of 12 cases indicated injuries at the para-sagittal area and base of the frontal and temporal lobe. Apart from these, four additional cases were published separately \cite{36ono1980human} and for which detailed autopsy results were available. Injuries reported for the four primates includes pulmonary hemorrhage, transection of medulla, hematoma in the pituitary gland, pulmonary congestion, cerebral hemorrhage and effect on medulla. The significantly injured area in these cases (apart from the area of impact) was near the base of the brain. The region effectively is farthest from the point of impact, which is predicted by the mechanics-based model. The super-ellipsoid closest fit to primate brain was utilized for pure tangential impact and results are presented in Figure \ref{tangent}, Figure \ref{9variation} and Figure \ref{10variation}. For the simulated case, the region of maximum trauma is farthest from the point of impact, as explained for the cases in section 3.
It was reported that for a similar impacting velocity, the primates survived during the near-normal impact but the tangential impact had fatal outcomes. This finding is consistent with the mechanics model, where  under near-normal impact, the predicted maximum energy of impact was around 27500 J whereas for the same wrench applied tangentially, the maximum energy of impact was around 58000 J. The ratio of the numbers for the peak energy for impacting wrench of the same magnitude was 1:2.1, suggesting higher intensity of energy and higher injury probability for tangential impact.
\subsubsection{General Impacting Wrench – Comparison to the experimental and finite element results}
The injuries resulting from a general wrench lateral impact are reported in \cite{37sakai1982experimental}. A combination of tangential and normal impact was applied with the head being impacted laterally resulting in a combination of linear and rotational acceleration of the head. Apart from the coup injuries (just below the point of impact), the injuries caused were primarily a combination of the near-normal impact and near-tangential impact. The major regions of injury included basal cistern and injury at the medullo-spinal junction, which can be explained by the tangential component of the impacting wrench, as these regions are farthest from the point of impact. Apart from this there were injuries in the parieto-occipital region as well, which can only be explained by the near-normal component of the impacting wrench, as the system is farthest from the line of impact.

The major injury is due to the tangential component usually but the near normal component also plays a role. As shown in Figure \ref{fig12}, there are two regions of local maxima of the energy based function, leading to two critical regions. A set of finite element simulations developed based on features of a Macaque monkey brain by the co-author and published in \cite{46antona2013correlation}\cite{47antona2012reanalysis}, report two non-coup injuries on the regions. The details of the FE simulations are published separately in \cite{46antona2013correlation}\cite{47antona2012reanalysis}. This can be attributed to the tangential component and normal component, when the general impacting wrench is discretized into a tangential component and a normal component. The result is shown schematically in Figure \ref{fig_jacobo14}.

\subsection{Results on True Brain Geometries}
The methodology is applied to brain of different shape and size. As, true brain geometries aren't regular, the instantaneous centers of curvature changes unlike symmetric super ellipsoids. We define near-normal impact as the impact passing closely to the geometric center of the geometry and near tangential as perpendicular to it. The method doesn't use the local curvature in defining near-normal or near tangential impacts. Also the closed form expression for the moments for true brain geometries are unknown, we calculate the integrals in equation \ref{moment1} numerically. Two sets of brain including a Macaque primate monkey brain and a Marmoset brain were simulated to find the most critical regions. \\
For the case of Macaque primate brain, which is similar to the human brain in shape various impact cases and subsequent injuries are presented in Figure \ref{comp_monkey15}. The variation of colour with the magnitude of injury is same as in figure \ref{tangent}. The top 10\% critical region is shown in the dark color. For all the cases in Figure \ref{comp_monkey15}, the arrow tail shows the point of impact except in Figure \ref{comp_monkey15} (c) where the arrow head shows the point of impact. The arrow shows the direction of impact. Figure \ref{comp_monkey15} (a) shows a tangential impact on the side of the brain, the most critical region being at maximum region from the point of impact. The most critical region occurs on the other side of the brain. Figure \ref{comp_monkey15} (b) shows an impact near the top of the brain, as a general impact, similar to that in Figure \ref{fig_jacobo14}, with critical regions resulting near to the occipital region of the brain. Figure \ref{comp_monkey15} (c) and Figure \ref{comp_monkey15} (d) are two cases of near normal impact, but with a slight variation in angle (around $20^o$). The most critical regions are primarily farthest from the line of impact. When the impact angle is directed more towards the spine, the occipital region starts coming into the critical regime. For a pure tangential impact (Figure \ref{comp_monkey15} (e)), the most critical region occurs at the back of the brain (occipital region).   
The marmoset brain differs from the human brain with the medial extent being greater than the vertical extent. A general impact to the marmoset brain was analysed for most critical regions. The point cloud for marmoset brain was generated using the CT images as in \cite{woodward2018brain}. The analysis results are presented in Figure \ref{marmoset}. The two views show the most critical region under a general impact which is more near normal than near tangent. The most critical area occurs on the temporal region as shown. 
\subsection{Comparison with other existent metric of concussion}
This section focuses on comparing the numerical values with existent metrics. The HIC value (equation \ref{HIC}) and the probability of concussion (equation \ref{PIC}) is compared for the super-ellipsoid approximation model. A force of 5000 N was applied for a duration of 0.02 s. The force was applied at the frontal surface with varying directions of impact. The idea is shown in the Figure \ref{show}. The direction of impacting force does not impact the HIC value as it is only affected by the translational acceleration but the probability of concussion shows correlation with the proposed metric. For the case in Figure \ref{show} the HIC value was constant at 5245 units. To compare the probability of concussion with the method proposed here, the energy terms are normalized by the maximum value (to compare with the probability of concussion). It can be observes that the method proposed here is coherent with the empirical relation suggesting probability of concussion.  

        
\subsection{Prediction of Impacting Wrench direction from human CT Data}
The methodology developed can also be used to predict the direction of the impact on the human head, knowing the position of the coup and the non-coup injury. The CT images of traumatic brain injury cases are shown in Figure \ref{fig:17}. In both the cases non-coup injury region is not furthest from the point of impact, ruling out a near tangential impact. We see that further away from the line of impact is a better descriptor than furthest from the point of impact. So, the major component of the impacting wrench would be normal to the surface.
Considering another case of a brain trauma case in JPNATC, AIIMS, Delhi, India, where the impact was at the frontal lobe and the CT image is presented in Figure \ref{fig:18}. As the non-coup region is farthest from the point of impact, it can be predicted that the impacting wrench would be primarily tangential. 
\noindent
\\
\\
\\
\\
\\
\\
\\
\\
\\
\\
\\
\\
\\
\noindent
\noindent
\noindent
\section{Conclusion}
This paper presents a method to predict the location of regions of brain trauma.  The method utilizes the direction of the impact force. Analytical estimates using an energy measure with a simplified brain geometry model indicates that human brain regions that sustain the highest damage potential are dependent on direction and position of impact. There is a large variation of the position of trauma between near normal, near tangential or a combination of these. These computations are consistent with past head impact experiments with non-human primates. The experimental observation that tangential impacts can be more injurious than corresponding normal impacts is consistent with this analysis. Further, the methodology proposed can also be applied to estimate the direction of the impact from the injury pattern, which should be useful to forensics.

%
%

\bibliographystyle{spmpsci}      


%
%
$\textbf{Appendix}$\\
$\textbf{A. Dynamics Using Screw Representations}$ \\
The general motion of a rigid body can be represented using the concept of motors. A motor ('mo' from moment and 'tor' from vector) is a combination of a vector and the moment of a vector. A general motor can also be represented as a complex vector \cite{twentyfive} with the real part as the vector and the imaginary part as the moment with respect to a point of observation (equation \ref{moment}).
\begin{equation}
    \mathbf{M} = \boldsymbol{\alpha}+i\boldsymbol{\beta}
\label{moment}
\end{equation}
where $\boldsymbol{\alpha}$ represents the real part and $\boldsymbol{\beta}$ represents the imaginary part. 
The total momentum of the body is represented by a momentum motor defined as $\mathbf{K (k, K_o})$ with real and imaginary vectors $\mathbf{k}$ and $\mathbf{K_o}$. To analyse the motion of the body following an impact, the kinematic motors and the force motors (wrenches) are related following the development in Dimentberg \cite{twentyfive}.
Dimentberg in particular, represents a general screw as a system of complex numbers. The instantaneous motion of the brain is related to the wrench $\mathbf{ W (F, M)}$ using the inertia binor matrix relating the instantaneous mechanical properties of the body. The inertia binor is represented as a rectangular matrix, with two $3\times3$ matrix with complex elements, separated by a comma (equation \ref{binor}). The methodology of representation of inertia binor as a combination of real and imaginary parts is useful as in the momentum equation we can directly compare the real and imaginary parts of the motor.
\newcommand\scalemath[2]{\scalebox{#1}{\mbox{\ensuremath{\displaystyle #2}}}}

\begin{equation}
    \textbf{T}=\left[\scalemath{0.9}{\begin{array}{ccccccc} iI_1& S_3-iD_3&-S_2-iD_2&m&-iS_3& iS_2&\\
    -S_3-iD_3&iI_2&-S_1-iD_1&iS_3&m& -iS_2&\\
    S_2-iD_2&-S_1-iD_1&iI_3&-iS_2& iS_1&m&
    \end{array}}\right]
    \label{binor}
\end{equation}

where ‘i’ as usual represents the imaginary part of a complex number and $S_i$ represents the first moment of inertia , $D_i$ are the cross product terms of the classical moment of inertia tensor and $I_i$ are the leading diagonal terms of the inertia tensor.    
The elements of the binor are a function of the spatial distribution of mass in the brain with respect to the point of observation (equation \ref{start_mom}-\ref{end_mom}). The total mass is m and the other elements of the binor are as per the equations \ref{start_mom}-\ref{end_mom} with $\rho_i, i = x, y, z$ being the respective Cartesian coordinates of a differential mass element dm. We note that the elements of the binor are a reassuring arrangement of classical mechanics inertia terms.
\begin{equation}
    \int\rho_xdm = S_1;   \int\rho_ydm = S_2;  \int\rho_zdm = S_3
    \label{start_mom}
\end{equation}

\begin{equation}
\scalemath{1}{
    \int\rho_y\rho_zdm = D_1;  \int\rho_x\rho_zdm = D_2; 
     \int\rho_y\rho_xdm = D_3}
\end{equation}
\begin{equation}
\scalemath{0.9}{
\int(\rho_x^2+\rho_y^2)dm = I_1;
    \int(\rho_x^2+\rho_z^2)dm = I_2; \\
    \int(\rho_z^2+\rho_y^2)dm = I_3}
    \label{end_mom}
\end{equation}
The dynamics of the body in the motor form reduces to a screw form differential equation (equation \ref{dyneq}) \cite{twentyfive}. The general solution of the differential equation provides the velocity motor with respect to the desired point of observation. Integration then provides the displacement motors.

\end{document}